# Fibre positioning revisited: the use of an off-the-shelf assembly robot for OPTIMOS-EVE

Gavin B. Dalton[*a,b], Martin S. Whalley[a], Oudayraj Mounissamy[b,c], Eric C. Sawyer[a], Ian A.J. Tosh[a], David L. Terrett[a], Ian J. Lewis[b].

[a]STFC-Rutherford-Appleton Laboratory, HSIC, Didcot, OX11 0QX, UK;
[b]Physics Department, University of Oxford, Keble Road, Oxford, OX1 3RH, UK;
[c]IUT Paris Jussieu, Unversité Paris Diderot, 2 Place Jussieu, case 7139, 75251 Paris, France.

**ABSTRACT**

The OPTIMOS-EVE instrument proposed for the E-ELT aims to use the maximum field of view available to the E-ELT in the limit of natural or ground-layer-corrected seeing for high multiplex fibre-fed multi-object spectroscopy in the visible and near-IR. At the bare nasmyth focus of the telescope, this field corresponds to a focal plane 2.3m in diameter, with a plate-scale of ~3mm/arcsec. The required positioning accuracy that is implied by seeing limited performance at this plate-scale brings the system into the range of performances of commercial off-the-shelf robots that are commonly used in industrial manufacturing processes. The cost-benefits that may be realized through such an approach must be offset against the robot performance, and the ease with which a useful software system can be implemented. We therefore investigate whether the use of such a system is indeed feasible for OPTIMOS-EVE, and the possibilities of extending this approach to other instruments that are currently in the planning stage.

## 1. INTRODUCTION

The use of optical fibres for astronomical multi-object spectroscopy was pioneered with the use the FOCAP plug-plate system at the Cassegrain focus of the AAT[1]. This basic approach proved to be extremely successful, and was employed by the Sloan-Digital Sky Survey to obtain spectroscopy of close to 1 million galaxies[2]. However, this approach is somewhat labour-intensive, and is unlikely to be acceptable within the environment of the E-ELT (although we note that a fully automated approach to fibre plug-plates has been suggested recently for the VLT[3]. The use of fibres was then developed into an automated positioner for the MX instrument[4], which used fibres mounted on deployable arms to obtain a modest multiplex with a low overhead at the telescope. This concept has recently been developed into a fully cryogenic system for KMOS at the VLT[5], but is generally of limited use in the high-multiplex domain due to the constraints of collisions between adjacent arms. Pick-and-place positioners, where a robot gantry is used to position magnetic fibre buttons within the focal plane, were developed as an upgrade to the FOCAP system at the AAT, with a progressive development though AUTOFIB[6], AF2[7] at the WHT, and finally the 2dF system at the AAT's prime focus[8]. Variants of these concepts, have been successfully deployed at the MMT[9] and the VLT[10]. These systems provide high multiplex and reasonable target proximity constraints, but are ultimately limited by the speed with which the fibres can be repositioned by a single robot gantry. Finally, we note here that a completely different local patrol field approach was developed to accommodate the small space envelope and high target densities required for the FMOS instrument on Subaru[11], which allows for high speed, independent positioning of a large number of fibres within a compact focal plane. This concept has been developed further[12,13,14], but has significant unit-cost implications.

---

[*]gavin.dalton@stfc.ac.uk; phone +44 1235 446401

## 2. FIBRE POSITIONING FOR OPTIMOS-EVE

The system design for OPTIMOS-EVE (EVE) is described in detail elsewhere in these proceedings[15,16]. Here we note only the main requirements of EVE that have an impact on the positioner concept design. EVE requires arbitrary targets located within the full 10' FOV of the E-ELT, and at f/17.8, this corresponds to a 2.16m diameter field with a concave radius of curvature of ~42m, and a plate-scale of 3.6mm/arcsec. For seeing limited apertures of 0.9" this implies a requirement on the positioning accuracy of 100-200μm, at least an order of magnitude more relaxed than the specifications for 2dF or FMOS. EVE also requires a number of different geometries for the input fibre bundles in order to obtain the required spectral resolutions and IFU modes. We initially considered employing a re-imaging focal reducer to provide a direct fibre input closer to f/4.5, with the aim of incorporating an atmospheric dispersion compensator within the corrector. This approach would produce a four-way split focal plane that would have been well-suited to a local patrol positioning scheme, but presents challenges in terms of the cost and mass of the required optics, and a large-non-telecentricity of the final focal plane. Furthermore, this approach does not appear well-suited to the multiple input modes required for EVE. We therefore adopted a pick-and-place solution for EVE using the bare focus of the telescope. The multiple input modes can be easily accommodated within such a system by means of a carousel system housing multiple field plates, and this also provides the option to duplicate those modes which require sufficiently high multiplex that the real-time reconfiguration time would become an unacceptable operational overhead.

### 2.1 Focal plane carousel

An initial approach to the carousel concept was derived from the VLT's existing FLAMES instrument, which allows for a gravity-stable positioner working upwards. While this approach is feasible, it requires most of the volume available to the entire instrument just to house the positioner assembly, since the active field plate must be mounted at the height of the Nasmyth optical axis. We therefore considered a more compact approach with four vertical field plates (see Figure 1). Since the telescope focal surface is ~750mm behind the published location of the adaptor rotator, this concept allows the carousel to rotate in situ, such that the volume swept out by the field plate as it comes into position is always clear of the rotator flange. The mass of each field plate (see Figure 2) is around 450kg, and so we propose to match the telescope field rotation with a rotator built into the carousel for each plate, rather than attempt to hand the plate off to a mounting on the adapter rotator. A significant advantage of this approach is that the entire carousel can be mounted on a platform that brings the field plates to the height of the Nasmyth axis, with the spectrographs accommodated underneath the positioner to give a relatively compact design and a short working fibre-length.

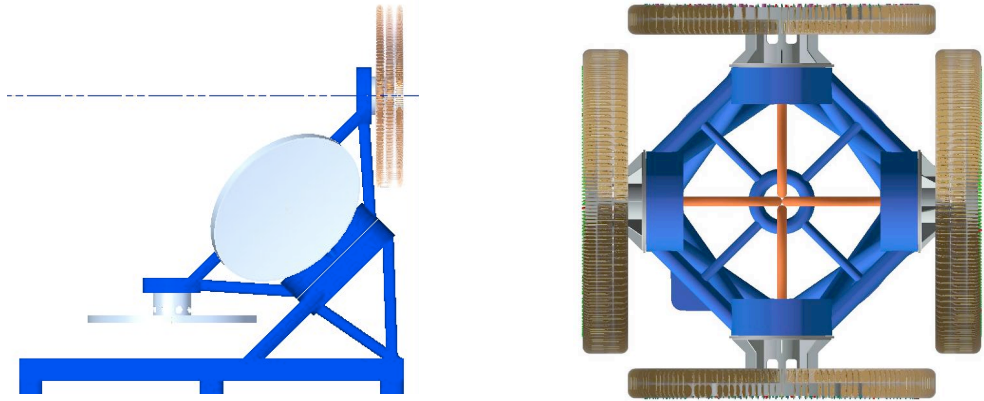

Figure 1. Two concepts considered for a positioner carousel. (left) Initial concept, similar to FLAMES with an upward-working positioner. (right) More compact concept with a sideways-working robot (viewed from above).

### 2.2 Fibre tensioners

Within a pick-and-place design, the fibre bundle for each button must be managed to ensure that the exit path of the fibre from the deployed button to the edge of the field remains straight. The compact concept for the carousel discussed above requires that the tensioners must be packed such that they fold underneath the field plate itself to avoid accommodation clashes with the adjacent plates. A simple consequence of this is that alternate tensioners around the field perimeter must extend to different depths beneath the field plate, as the space available close to the centre of the volume is much less than that around the perimeter (Figure 2). On the other hand, this geometry provides a natural exit for the fibres towards the hub of the field plate assembly, and hence a short exit path through the centre of the carousel and down towards the spectrographs.

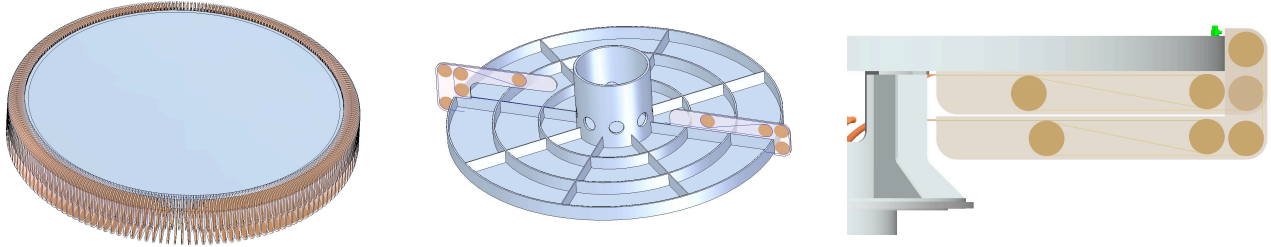

Figure 2. (left) Field plate showing 240 individual fibre retractors. (center) Backside of the plate showing stiffening structure and the two different retractor geometries. (right) Side view of two adjacent retractors showing accommodation beneath the plate.

## 2.3 Availability of Commercial Robots

We considered both XYZ gantry systems and articulated assembly robots for OPTIMOS-EVE. The results of an initial set of queries suggested that the gantry systems could be difficult to implement efficiently in such a large vertical configuration, and so the main body of this work has concentrated on articulated systems (more recent discussions with Schunk imply that gantry-systems may indeed be applicable in this case, but are more likely to be applicable to smaller scale problems with more stringent positioning requirements, such as GYES[17] and the WHT-MOS[18]). In either case, the motion speed of the robots is likely to far exceed the operational speed that can be used without causing failure of the fibre tensioners.

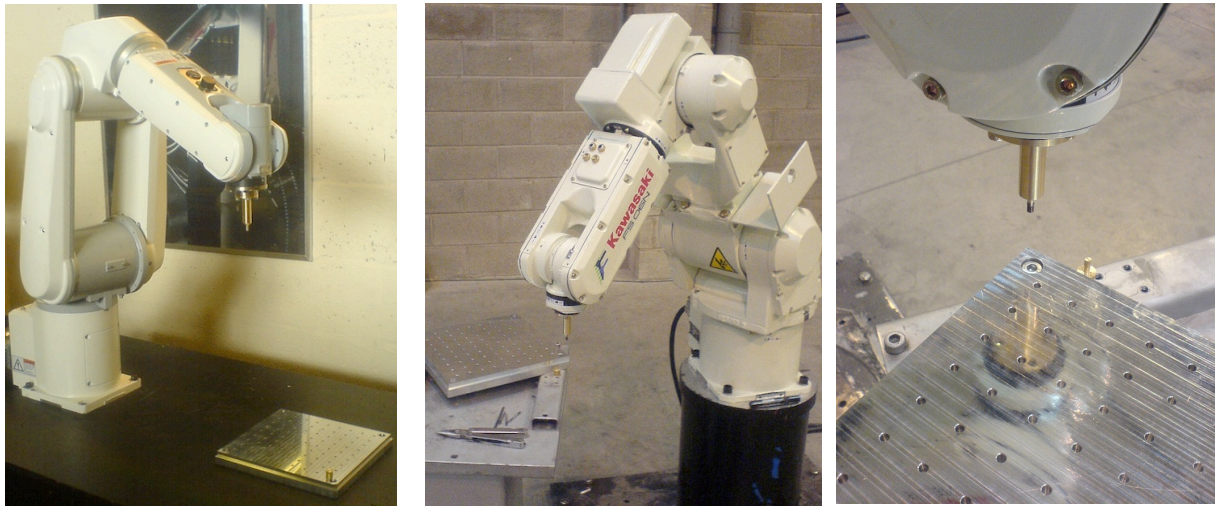

Figure 3: Initial factory testing of articulated assembly robots (left: Toshiba; center: Kawasaki; right: closeup of the test plate and pin).

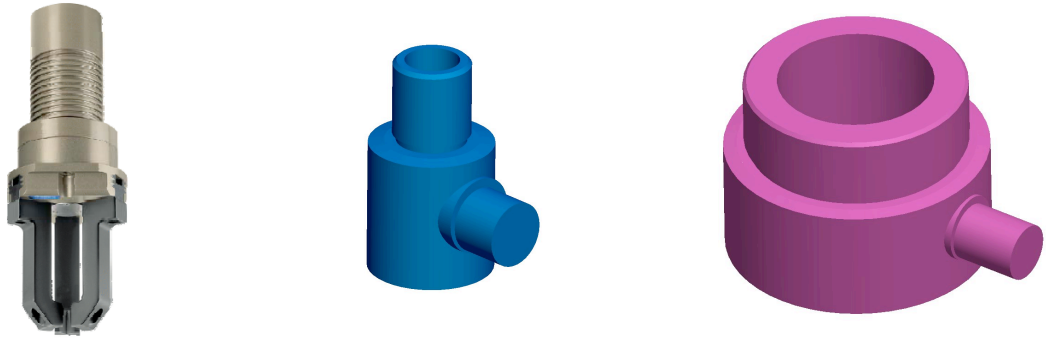

Figure 4: 3-Finger gripper unit (Schunk) used in the pick-and-place trials (left), together with geometrical layouts for the multi-object and multi-IFU buttons used in the OPTIMOS-EVE design.

We contacted the UK distributors for Toshiba Robotics and Kawasaki Robots to arrange for a preliminary assessment of possible systems. A potential issue for these systems appeared to be that the manufacturers quote positioning accuracies that are strictly only applicable to the manufacturing case of a systematic return to pre-taught pose positions. We therefore constructed a simple test consisting of a 240x240mm regular plug-board with 3.2mm diameter holes on a 20mm pitch and a 3mm tapered pin that could be attached in place of the robot 'hand' (Figure 3). The robot may then pick up the co-ordinate reference frame from the corner holes of the plate and attempt to insert the pin into each hole in turn as a previously undetermined position. While this test is a substantial oversimplification of the true positioning problem, it nevertheless provides a useful pass/fail test to decide whether or not to proceed with this avenue of exploration. The Kawasaki robot performed best in these early trials, and was adopted as a baseline solution for OPTIMOS-EVE. We subsequently acquired a smaller version of this robot for more detailed investigation of the operational aspects of the software interface, and to extend the test to open-loop pick-and-place testing using an off-the-shelf gripper unit supplied by Schunk (Figure 4).

## 3. OPTIMOS-EVE POSITIONER DESIGN

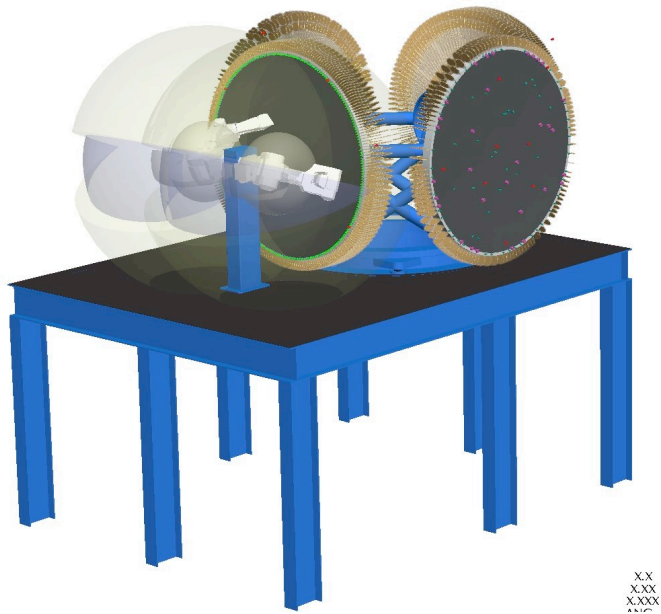

Figure 5: OPTIMOS-EVE carousel mounted on the instrument platform with two robots working on the configuration.

Although it appears possible to design OPTIMOS-EVE with a single robot, the limitations on positioning speed that are expected to arise from the tensioner design implies that two robots working in tandem will be required to achieve a field reconfiguration time of less than 30 minutes. The layout of the instrument with the swept working volumes for the two robots is shown in Figure 5. The OPTIMOS-EVE system design[15] allows for the fibres that are being configured to be illuminated from within the spectrographs, and there is ample space for a small imaging camera to be mounted within the jaws of the gripper unit shown in Figure 4. Our goal is to arrive at a system that will deliver a positional accuracy of < 150µm in open loop, but with the imaging system required for initial calibration and long-term quality control monitoring. The carousel arrangement described in Section 2 allows for a relatively simple solution for the management of the fibre cables between the field plates and the spectrographs (Figure 6).

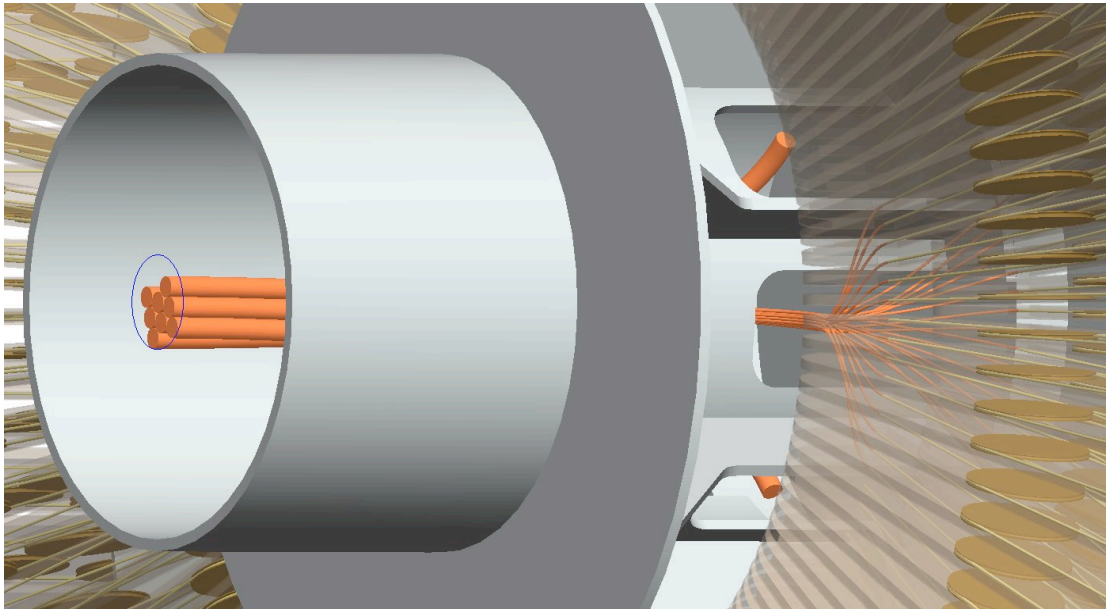

Figure 6: Routing of the fibre cables from the fixed point of each tensioner unit through the axis of the plate rotation bearing. From this point the fibres pass down through the center of the carousel rotation bearing, through the positioner platform and into the spectrograph enclosures.

We have simulated the configuration of OPTIMOS-EVE fibres using the button geometries shown in Figure 4 and a development of the configuration scheme used to define the 2dF Galaxy Redshift Survey[19]. The software allows for different types of buttons to be configured within the same field, allowing for medium resolution single objects and multi-IFU configurations to be deployed on the same field plate at the start of the night, as shown in the side-position field plate in Figure 5. The configuration allows for fibre bundles to cross on the field plate, but forbids collision of the buttons with buttons or deployed fibres. Excluded areas around the field perimeter due to the vignetting footprints of the telescope wavefront sensor pick-offs have not yet been included.

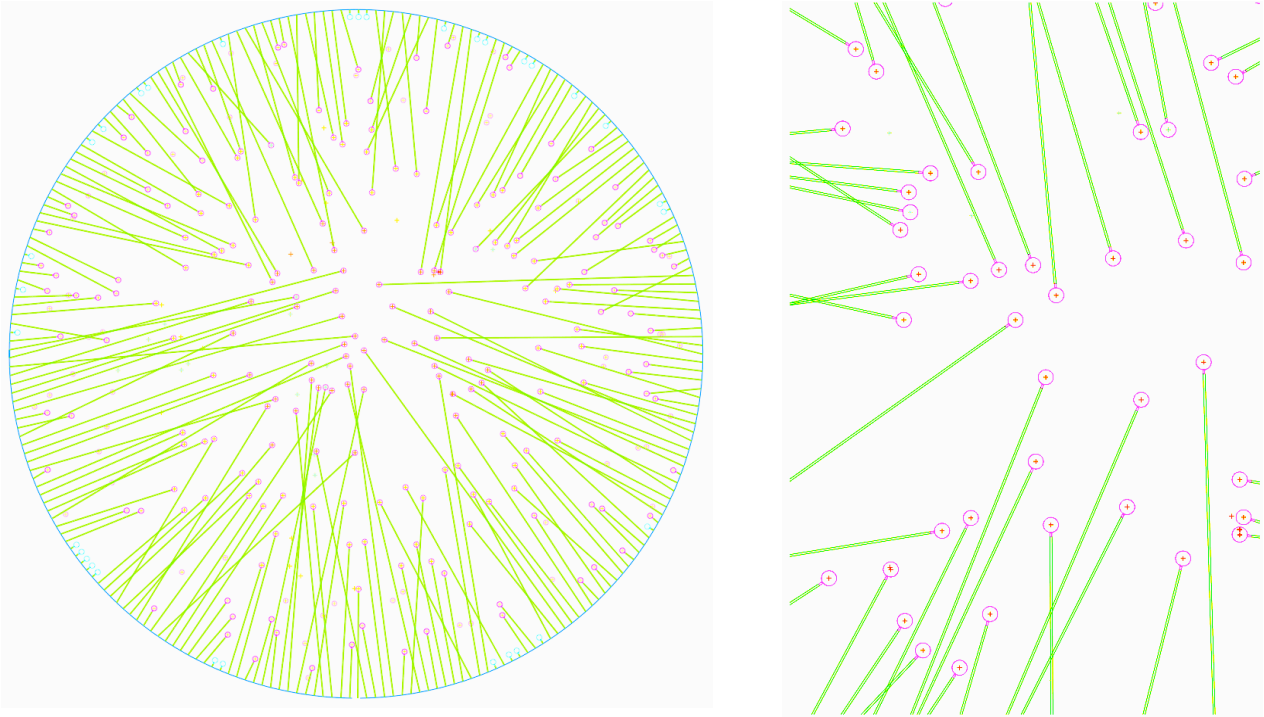

Figure 7: Sample configured field and closeup with 245 16mm diameter OPTIMOS-EVE buttons on a 10' (2.2m) diameter focal plane. 209 fibres have been successfully deployed on a random distribution of targets in this example.

## 4. LABORATORY TEST SETUP

The testing of the robot performance for a real pick-and-place setup is underway in Oxford. For the first phase, we have constructed mechanical dummy EVE buttons using rare-earth magnets embedded in steel buttons. These can be placed by the robot at specified locations on a 0.6x1.2m steel plate within a coordinate reference frame defined by three buttons placed close to the field corners. Once the buttons are positioned, the plate can be transferred directly to our coordinate measuring machine, which references each position to the same internal coordinate frame. The configuration of the test setup allows tests to be carried out in horizontal and vertical arrangements. The primary aim of this test is to allow the implementation of a set of open loop correction to the robot's motion to correct for any systematic inaccuracies in the robot itself, and also to compensate for differences in the pick-up and put-down behaviour of the individual buttons.

Once this initial phase is complete the setup can easily be extended to include optical measurements from fibre buttons, and to test the reliability, speed, and impact on the positioning performance of representative tensioner units. The setup will also be used to investigate the vibration spectrum that could be transmitted to the instrument from the robot. With the inclusion of a modest size alt-azimuth telescope mount available to us, this setup can also easily be adapted to tests for prime-focus pick-and-place solutions, such as those being considered for GYES on the CFHT, and the WHT-MOS system.

## 5. SUMMARY

We have presented a design solution for the OPTIMOS-EVE fibre positioner that uses commercial assembly robots to minimize the cost, effort and risk associated with the development of a robust and efficient system for the E-ELT. Initial tests conducted with the suppliers suggest that these systems are indeed capable of delivering the performance required for a fibre positioner working at the E-ELT Nasmyth focus. A simple test platform has been set up in Oxford to allow step-by-step performance testing and development of this concept.